\documentclass[preprint]{jfm}

\usepackage{graphicx}
\usepackage{amsmath,amssymb}
\usepackage{natbib}
\usepackage{hyperref}
\usepackage{listings}
\hypersetup{
    colorlinks = true,
    urlcolor   = blue,
    citecolor  = black,
}

\newcommand{\RomanNumeralCaps}[1]
\linenumbers

\newcommand\bh{\mathbf{h}}
\newcommand\bH{\mathbf{H}}
\newcommand\bk{\mathbf{k}}
\newcommand\bp{\mathbf{p}}
\newcommand\bq{\mathbf{q}}

\newcommand\bu{\mathbf{u}}
\newcommand\bv{\mathbf{v}}
\newcommand\bua{\mathbf{u}_{\rm a}}
\newcommand\bva{\mathbf{v}_{\rm a}}
\newcommand\bug{\mathbf{u}_{\rm g}}
\newcommand\bvg{\mathbf{v}_{\rm g}}
\newcommand\bw{\boldsymbol{\omega}}
\newcommand\bpsi{\boldsymbol{\psi}}
\newcommand\bX{\mathbf{X}}

\title{Quasi-geostrophic projection of rotating shallow-water equations}

\author{
  Louis Thiry\aff{1,2},
    \corresp{\email{louis[dot]thiry[at]inria[dot]fr}}
  Long Li\aff{1},
  Etienne Mémin\aff{1},
  \and Guillaume Roullet\aff{3}
}
\affiliation{
  \aff{1}Equipe ODYSSEY, INRIA Rennes, IRMAR Rennes, Université Rennes 1, Rennes, France
  \aff{2}Equipe ANGE, INRIA Paris, Laboratoire Jacques-Louis Lions, Sorbonne Universite, Rennes, France
  \aff{3}Université de Bretagne Occidentale , CNRS, IRD, Ifremer, Laboratoire d'Océanographie Physique et Spatiale (LOPS), IUEM, Brest, France
}

\begin{document}
\maketitle

\begin{abstract}
With a new unifying model for layered rotating shallow-water (RSW) and quasi-geostrophic (QG) equations, this paper sheds light on the relation between these two sets of equations.
We propose here a new formulation of the quasi-geostrophic equations as a projection of the rotating shallow-water equations.
This QG formulation uses the same prognostic variables as RSW, namely velocity and layer thickness, restoring the proximity of these two sets of equations.
It allows direct access to the ageostrophic velocities hidden in geostrophic velocities resolved by the QG equations.
It opens the path of studying the difference between QG and RSW using the same underlying numerical discretization.
We illustrate this new possibility on a vortex shear instability and a double-gyre configuration.
\end{abstract}



\section{Introduction}

Large-scale ocean models offers a natural trade-off between complexity and realism. On the one side, the primitive equations (PE) are realistic enough to be used for climate simulations and real-world data assimilation. However, they are rather complex as they describe the coupling between equation of motions (mass and momentum) and conservation of tracers via the equation of state following thermodynamics laws. On the other side, the barotropic planetary geostrophic equations can be efficiently solved with a single prognostic variable (the thickness), yet they ignore thermodynamics and vertical variations.

In between, there exists some approximate models of the PE such as the multi-layer rotating shallow-water (RSW) and the multi-layer quasi-geostrophic (QG) models.
The latter can be either derived from the former using an asymptotic approach \citep{pedlosky2013geophysical,vallis2017atmospheric} or considered as a vertical discretization of the continuously stratified QG model derived from the PE.
These layered models describe the dynamics of vertically stratified flow in isentropic (or isopycnal) coordinates and only require solving the horizontal momentum and mass equations.
For instance, the multi-layer RSW model can fairly reproduce the ocean dynamics in the Gulf-Stream region with solely five vertical levels \citep{hurlburt2000impact}.
The multi-layer QG equations are widely adopted for the development of determistic mesoscale eddies parameterizations \citep{marshall2012framework,jansen2014parameterizing, fox2014principles,ryzhov2020data,uchida2022deterministic} as well as stochastic ones \citep{grooms2015stochastic, zanna2017scale,bauer2020stochastic}.
These two sets of equations can be used as research tools as their numerical integration is light enough to run on laptop computers.

Although the multi-layer QG equations are derived from the multi-layer RSW equations, their relationship is still not clear from a numerical point of view.
Indeed, they are usually written with different progonstic variables.
The prognostic variables of the RSW system are the horizontal velocity $(\bu, \bv)$ and the layer thickness $\bh$, whereas one can formulate the QG system with a single prognostic variable, typically the potential vorticity $\bq$ from which one can diagnose the streamfunction $\bpsi$ and the pressure $\bp$.
This difference breaks the conceptual continuity between these two equation sets in the ocean model hierarchy.
It also brings some undesirable consequences in practice.
For instance, many eddy parameterizations \cite[e.g.][]{bachman2019gm+,li2023stochastic} yield different formulations and/or discretizations when applied to the RSW model, using $(\bu,\bv,\bh)$ as the prognostic variables, or applied to the QG model, using $\bq$ as the prognostic variable.
Moreover, it is not straightforward to use the same discrete schemes for these two different models in order to compare them under the same configuration.

In the present work, we propose to reformulate the multi-layer QG model using $(\bu, \bv, \bh)$ as the prognostic variable.
The QG equations can be written as a projection acting on the RSW equations.
This is the main novelty of our paper.
Using this projection approach, one can build a QG discretization on top of any RSW discretization. Moreover, this formulation provides direct access to the ageostrophic velocity that is hidden in the standard QG model formulation, yet contributes to QG dynamics.

For the numerical experiments, we adopt the RSW discretization proposed by \cite{roullet2022fast} which relies on the vector invariant formulation and uses high-order WENO reconstructions to provide implicit dissipation.
The discretized QG model uses the same dynamical core than the multi-layer RSW model. The only modification is to add the projection operator.
We end up with a compact, efficient, and CPU/GPU portable Python code using the PyTorch library \citep{paszke2019pytorch}.
We first use a test case of vortex shear instability to investigate the similarities and differences of the QG and RSW solutions according to different Rossby numbers.
We then study the solutions produced by QG and RSW on an idealized double-gyre configuration.

This paper is organized as follows.
In Section 2, we briefly recall the derivation of the multi-layer RSW and QG equations.
In Section 3, we present our projected QG formulation.
In Section 4, we numerically test our new formulation on two different configurations.
We conclude and evoke further perspectives in Section 5.

\section{Multi-layer RSW and QG equations}

In this section, we first review the governing equations of the RSW system, then explain briefly the QG scaling of RSW equation and give the usual multi-layer QG equations using $\bq$ as the prognostic variable.

\subsection{Multi-layer RSW equations}

\begin{figure}
    \centering
    \includegraphics[width=0.8\linewidth]{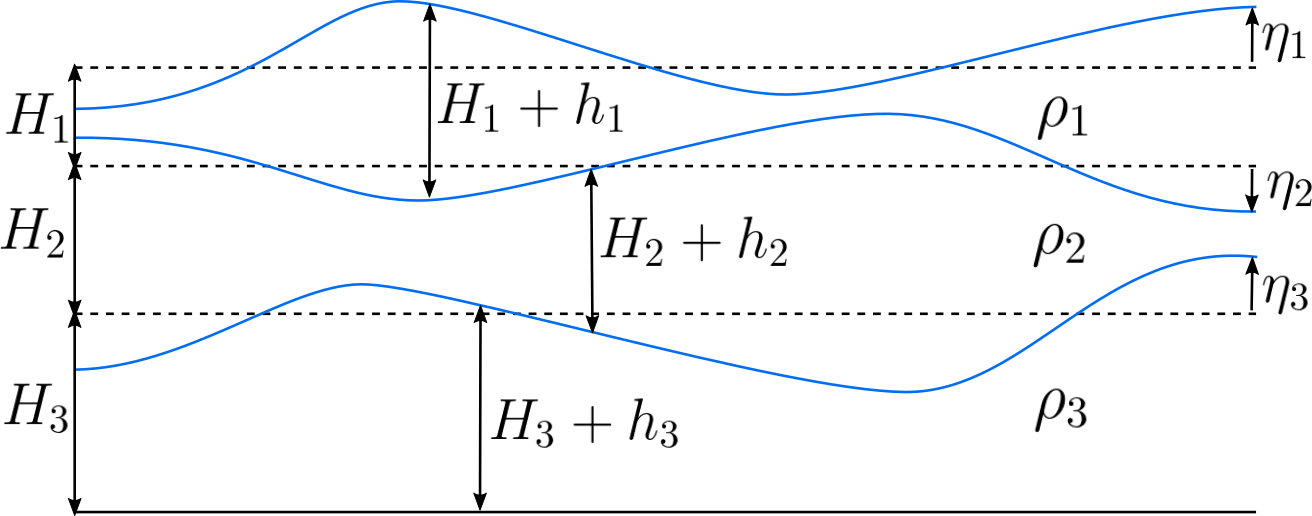}
    \caption{Vertical cross-section of a three-layer RSW model displaying layer thicknesses $H_i + h_i$ and interface heights $\eta_i$ }
    \label{fig:sw}
\end{figure}

The stratification of a multilayer RSW model consists of a stack of $n$ isopycnal layers, as illustrated in Figure \ref{fig:sw} with $n=3$ layers. By convention we index the layers from $i=1$ for the top layer to $i=n$ for the bottom one.
These layers have a uniform reference thickness $H_i$ and a density $\rho_i$.
As shown in Figure \ref{fig:sw}, the total thickness of a layer $i$ is the sum of the reference thickness $H_i$ and the thickness anomaly $h_i(x,y)$.
The interface vertical displacment $\eta_i(x,y)$ and the hydrostatic pressure $p_i(x,y)$ are given by
\begin{equation}
  \eta_i = \sum_{j=i}^{n} h_j \ , \quad
  p_i = \rho_1 \sum_{j=1}^i g_j' \eta_j
\end{equation}
with, for $i>1$, the reduced gravities $g'_{i} = g (\rho_{i} - \rho_{i-1}) / \rho_1$ and for the top layer $g'_1 = g$.

Using the following vector notation
\begin{subequations}
\begin{align}
   \bH & = \left( H_1,\  \ldots,\ H_n \right), \\
  \bh &= \left( h_1(x,y),\  \ldots,\  h_n(x,y) \right), \\
  \boldsymbol{\eta} &= \left( \eta_1(x,y),\  \ldots,\  \eta_n(x,y) \right), \\
  \bp &= \left( p_1(x,y),\  \ldots,\  p_n(x,y) \right),
\end{align}
\end{subequations}
we can rewrite the linear relations between $\bh, \boldsymbol{\eta}$ and $\bp$ in a compact form
\begin{subequations}
\begin{align}
    \bh & = \mathop{diag}(\bH) A \bp , \label{eq:hydro_hp} \\
    \boldsymbol{\eta} & =  C \bh \label{eq:eta_h} \\
    \bp & = M \bh, \label{eq:psw_h}
\end{align}
with the matrices
\begin{align} 
    A & = \frac{1}{\rho_1}
        \begin{bmatrix}
        \frac{1}{H_1 g_1'} + \frac{1}{H_1 g_2'} & \frac{-1}{H_1 g_2'} & . & . \\
        \frac{-1}{H_2 g_2'} & \frac{1}{H_2}\left( \frac{1}{g_2'} + \frac{1}{g_3'}\right)  &  \frac{-1}{H_2 g_3'} & . \\
        . & \ldots & \ldots & \ldots \\
        . & . &  \frac{-1}{H_n g_n'} & \frac{1}{H_n g_n'}\\
        \end{bmatrix} \ \label{eq:strechop}, \\
    B &=
    \rho_1
     \begin{bmatrix}
        g'_1 & 0    & 0    & \ldots & 0  \\
        g'_1 & g'_2 & 0    & \ldots & 0  \\
        .    & .    & .    & .      & . \\
        g'_1 & g'_2 & g'_3 & \ldots &  g'_n
    \end{bmatrix}, \ \
    C =
     \begin{bmatrix}
        1 & 1 & \ldots & 1 & 1  \\
        0 & 1 & \ldots & 1 & 1  \\
        . & . &  .     & . & .  \\
        0 & 0 & \ldots & 0 & 1
      \end{bmatrix},  \label{eq:BC_matrices}
\end{align}
\end{subequations}
and $M=BC$.
We then introduce the velocity $(\bu, \bv)$, the kinetic energy $\bk$, and the relative vorticity $\bw$
\begin{subequations}
\begin{align}
  \bu &= \left( u_1(x,y), \ldots, u_n(x,y) \right), \\
  \bv &= \left( v_1(x,y), \ldots, v_n(x,y) \right), \\
  \bk &= (\bu^2 + \bv^2) / 2, \\
  \bw &= \partial_x \bv - \partial_y \bu \ .
\end{align}
\end{subequations}
With these variables, the multi-layer RSW equations read
\begin{subequations}
\begin{align}
  &\partial_t \, \bu = (\bw + f) \bv - \partial_x (\bp + \bk) , \label{eq:rswmu}\\
  &\partial_t \, \bv = -(\bw + f) \bu - \partial_y (\bp + \bk) , \label{eq:rswmv}\\
  &\partial_t \, \bh = - \bH (\partial_x \bu + \partial_y \bv) - \partial_x (\bu  \bh) - \partial_y (\bv \bh) \,, \label{eq:rswmh}
\end{align}
\end{subequations}
where $f$ is the Coriolis parameter.
They can be formulated in the compact form
\begin{subequations}
\begin{equation}
    \partial_t \bX = F(\bX) \,, \label{eq:rswvec}
  \end{equation}
  with $\bX = \left(\bu, \bv, \bh \right)^T$, the state variable, and 
\begin{equation}
  \label{eq:rswvec_rhs}
  F
  \begin{pmatrix}
  \bu\\
  \bv\\
  \bh
  \end{pmatrix}
  =
  \begin{pmatrix}
    (\partial_x \bv - \partial_y \bu + f) \bv - \partial_x \left( M \bh + (\bu^2 +\bv^2) / 2 \right) \\
    -(\partial_x \bv - \partial_y \bu  + f) \bu - \partial_y \left( M \bh + (\bu^2 + \bv^2) / 2 \right) \\
    - \bH(\partial_x \bu + \partial_y \bv) - \partial_x (\bu \bh) - \partial_y (\bv \bh) 
  \end{pmatrix}
  \,,
\end{equation}
the RSW model operator.
\end{subequations}

\subsection{QG scaling of RSW equations}
\label{ssec:2.2}
The QG model is derived from the RSW model under two scaling assumptions: $Ro\ll 1$ and  $Bu\sim 1$ \citep{mcwilliams2006fundamentals,zeitlin2018geophysical} where $Ro$ is the Rossby number and $Bu$ the Burger number. The consistency of the QG scaling imposes a beta-plane approximation for the Coriolis parameter $f = f_0 + \beta y$, where $\beta$ is the meridional Coriolis parameter gradient. $Ro\ll 1$ implies that the velocity is close to the geostrophic balance, and specifically that the ageostrophic correction is $O(Ro)$. We define $(\bug, \bvg)$, the geostrophic velocity that obeys the geostrophic balance
\begin{subequations}
\begin{align}\label{eq:gbuv}
    -f_0 \bug = \partial_y \bp,\ \quad f_0 \bvg = \partial_x \bp \,,
\end{align}
and decompose the velocity into the geostrophic and the ageostrophic velocity $(\bua, \bva)$
\begin{align}
\label{eq:ageos_corr}
    \bu = \bug + \bua , \quad \bv = \bvg + \bva \,.
\end{align}
\end{subequations}
$Bu\sim 1$ and $Ro\ll 1$ imply that $\bh/\bH\sim O(Ro)$.
Following the approach of \cite{holton1973introduction,mcwilliams2006fundamentals,cushman2011introduction,zeitlin2018geophysical},
we do not decompose $\bp, \bh, \boldsymbol{\eta}$ in a  geostrophic and an ageostrophic part. This would require an extra assumption to decide what is the ageostrophic part of the stratification. This is unecessary to derive the QG model. There is therefore a single mass variable and two velocities per layer.

With these assumptions, the multi-layer QG equations can be written in terms of three equations for the prognostic variables $(\bug, \bvg, \bh)$ with the right-hand side terms that depend on the ageostrophic velocities $(\bua, \bva)$
\begin{subequations}
\begin{align}
  &\partial_t \, \bu_g = (\bw_g +  f_0 + \beta y) \bv_g + f_0 \bv_a - \partial_x (M \bh + \bk_g),  \label{eq:qgswu}\\
  &\partial_t \, \bv_g = -(\bw_g + f_0 + \beta y) \bu_g - f_0 \bu_a - \partial_y (M \bh + \bk_g) \label{eq:qgswv},\\
  &\partial_t \, \bh =  - \bH (\partial_x \bu_g + \partial_y \bv_g)  -\partial_x (\bu_g \bh) - \partial_y (\bv_g \bh) - \bH (\partial_x \bu_a + \partial_y \bv_a) \label{eq:qgswh}.
\end{align}
\end{subequations}
Note that using the geostrophic balance \eqref{eq:gbuv}, one can simplify the above equations into
\begin{subequations}
\begin{align}
  &\partial_t \, \bu_g = (\bw_g +  \beta y) \bv_g + f_0 \bv_a - \partial_x \bk_g \label{eq:qgswu_s} \ , \\
  &\partial_t \, \bv_g = -(\bw_g + \beta y) \bu_g - f_0 \bu_a - \partial_y \bk_g \label{eq:qgswv_s} \ , \\
  &\partial_t \, \bh = -\partial_x (\bu_g \bh) - \partial_y (\bv_g \bh) - \bH (\partial_x \bu_a + \partial_y \bv_a) \label{eq:qgswh_s}.
\end{align}
\end{subequations}
In this form the QG equations are still in appearance very close to the RSW equations, but this is misleading for there is not three degrees of freedom per layer, but only one. The two components of the geostrophic velocity are slaved to $\bh$ via \eqref{eq:gbuv} and \eqref{eq:hydro_hp}. Another difficulty with this system is that, though well posed, it is highly implicit. There is no obvious way to see how to determine the ageostrophic velocity. In practice, and in particular for numerical models, the model equations are rewritten in a more explicit form. This form reveals a hidden variable of the system: the potential vorticity (PV)
\begin{equation}
  \bq = f_0+\beta\,y + \bw_g - f_0\frac{\bh}{\bH} \,.\label{eq:pvdef}
\end{equation}

\subsection{Multi-layer QG equations}
The multi-layer QG equations are obtained upon derivating in time \eqref{eq:pvdef} and using \eqref{eq:qgswu_s}-\eqref{eq:qgswh_s} 
\begin{subequations}
\begin{align}
  & \partial_t \, \bq = -\partial_x (\bug \bq) - \partial_y (\bvg \bq ) , \label{eq:qgmadv}\\
  & \Delta \bp - f_0^2 A \bp = f_0 \bq - f_0 \beta y ,                        \label{eq:qgmell}\\
  & -f_0 \bug = \partial_y \bp,\ \quad f_0 \bvg = \partial_x \bp                \label{eq:qgmgeo} ,
\end{align}
\end{subequations}
where $\Delta = \partial_{xx}^2 + \partial_{yy}^2$ denotes the horizontal Laplacian operator and $A$ is the vertical discretization of the stretching operator introduced in Equation \eqref{eq:strechop}. The PV $\bq$ is the sole prognostic variable \eqref{eq:qgmadv} of the model. All the other model variables are deduced from it via the diagnostic relations \eqref{eq:qgmell} for $\bp$ and \eqref{eq:qgmgeo} for $\bu$ and $\bv$. The QG model is written in $(\bq,\,\bp)$ variables and $\bp/f_0$ is the vector of stream-function for each layer. Numerical implementations of the QG model usually rely on this system of equations. However, this system is now very different from the RSW system of equations. One aspect is the dispareance of $\bua$ from the equations. This does not mean that $\bua$ and $\bva$ vanish, rather, it means that $\bua$ and $\bva$ can be ignored if one is interested only in the evolution of $\bp$ and $\bq$. The large difference between the model equations  makes the comparison between them harder. In particular, and because of that, the numerical versions of QG and RSW  models generally have completely different implementations, with different numerics on all aspects. We now show how to restore the proximity between the two.

\section{Quasi-geostrophic model as projected rotating shallow-water}

In this section, we propose a new formulation of the QG equation using $(\bu, \bv, \bh)$ as state variables, the non-linear RSW operator $F$ defined in  \eqref{eq:rswvec_rhs}, and a linear projection operator $P$.
Strikingly, to our knowledge, this projection relation has been mostly under the radar except for one mention in \citep{charve2004etude}.

\subsection{QG as projected RSW}
To derive this projection, we start from the QG equations \eqref{eq:qgswu}, \eqref{eq:qgswv}, and \eqref{eq:qgswh}, and isolate the part controlled by $F$
\begin{equation}
\label{eq:F_plus_ag}
\partial_t
  \left(
  \begin{matrix}
    \bu_g\\
    \bv_g\\
    \bh
  \end{matrix}
\right)
  =
  F\left(
  \begin{matrix}
    \bu_g\\
    \bv_g\\
    \bh
  \end{matrix}
  \right)
  +
  \left(
  \begin{matrix}
     f_0 \bv_a\\
    -f_0 \bu_a\\
    - \bH (\partial_x \bu_a + \partial_y \bv_a)
  \end{matrix}
  \right),
\end{equation}
in which several terms cancel due to  the geostrophic balance.
In the form \eqref{eq:F_plus_ag} the QG model can be interpreted as a RSW model forced by an ageostrophic  source term, the second term of the right-hand side. Let us now introduce the PV linear operator $Q$
\begin{subequations}
\begin{equation}
Q \left(
    \begin{matrix}
    \bu\\
    \bv\\
    \bh
    \end{matrix}
  \right) = \partial_x \bv - \partial_y \bu - f_0 \frac{\bh}{\bH} \,.
\end{equation}
This operator is related to the PV $\bq$ by
\begin{equation}
Q \left(
    \begin{matrix}
    \bu_g\\
    \bv_g\\
    \bh
    \end{matrix}
  \right) = \bq - \beta y 
\end{equation}
and makes the contribution of the ageostrophic source term in \eqref{eq:F_plus_ag} vanishes 
\begin{equation}
  Q \left(
  \begin{matrix}
     f_0 \bv_a\\
    -f_0 \bu_a\\
    - \bH (\partial_x \bu_a + \partial_y \bv_a)
  \end{matrix}
  \right)
  =
  0.
\end{equation}
\end{subequations}
Since $Q$ is independant of time,  it commutes with the time derivative, therefore
\begin{equation}
\partial_t
  Q \left(
  \begin{matrix}
    \bu_g\\
    \bv_g\\
    \bh
  \end{matrix}
\right)
  =
  Q \circ F\left(
  \begin{matrix}
    \bu_g\\
    \bv_g\\
    \bh
  \end{matrix}
  \right) .
\end{equation}
This equation is simply a reformulation of \eqref{eq:qgmadv}, the conservation of PV. Notice that because $Q$ decreases the size of the state vector from three to one variable, we need a reverse operation to retrieve the three variables $(\bu_g, \bv_g, \bh)$. This is done by introducing the geostrophic  operator $G$
\begin{subequations}
\begin{equation}
  G\left( \bp \right) =
  \begin{pmatrix}
      - \partial_y \bp  \\
      \partial_x \bp \\
      f_0\,\bH A \bp
  \end{pmatrix} \,.
\end{equation}
$G$ is a linear operator with uniform and time independant
coefficients. The composition of the two operators $Q$ and $G$ results
in
\begin{equation}
  (Q \circ  G) \left( \bp \right) = \Delta \bp - f_0^2 A \bp \,
\end{equation}
\end{subequations}
which is the QG elliptic operator \eqref{eq:qgmell} that relates $\bp$ to $\bq$. This operator is invertible \citep{dutton1974nonlinear,bourgeois1994validity}, therefore $(Q \circ  G)^{-1}$ is well defined.
Finally, we introduce the QG projection operator $P$
\begin{align}
P   = G \circ (Q \circ G) ^{-1} \circ Q \, .
\label{eq:qg_projector}
\end{align}
$P$ is a projection because $P\circ P=P$, which is easily proven
\begin{align}
P \circ P = G \circ (Q \circ G) ^{-1} \circ (Q  \circ G) \circ (Q \circ G) ^{-1} \circ Q = P \,.
\end{align}
By construction, $P$ preserves the geostrophic state  $\bX_g = \left(\bu_g, \bv_g, \bh\right)^T$, namely $ P(\bX_g) = \bX_g$ .
Indeed, since $\bX_g$ is geostrophic, we have
\begin{align}
    Q(\bX_g) &=  \bq-\beta\,y \,,\\
  (Q \circ G) ^{-1}(\bq-\beta\,y) &= \bp/f_0 \,,\\
  G(\bp/f_0) &= \bX_g \,.
\end{align}
By applying $P$ on \eqref{eq:F_plus_ag}, and noting that $P$ commutes with $\partial_t$, we can formulate the multi-layer QG model in the form
\begin{equation}
\label{eq:qg_projected_sw}
\partial_t \bX_g  = P \circ F\left( \bX_g \right)  \,.
\end{equation}
This form differs from the RSW model \eqref{eq:rswvec} by the additional projection operator $P$ acting on $F$. By projecting the RSW tendency on the geostrophic manifold, $P$ ensures that the QG state remains in geostrophic balance. Another way to express it, is to say that the QG model evolves under the action of the RSW operator in which the ageostrophic tendency is removed.

\subsection{Ageostrophic velocities}

With this formulation, the ageostrophic velocity $(\bu_a, \bv_a)$ has a simple closed form.
Using equations \eqref{eq:F_plus_ag} and \eqref{eq:qg_projected_sw}, we have
\begin{equation}
\label{}
\left(
\begin{matrix}
    f_0 \bv_a\\
  -f_0 \bu_a\\
  - \bH (\partial_x \bu_a + \partial_y \bv_a)
\end{matrix}
\right)
=
P \circ F
  \left(
  \begin{matrix}
    \bu_g\\
    \bv_g\\
    \bh
  \end{matrix}
\right)
-
F\left(
\begin{matrix}
  \bu_g\\
  \bv_g\\
  \bh
\end{matrix}
\right) \, .
\end{equation}
In a numerical model this requires no costly computations, only a basic difference between the QG and the RSW tendencies.

\subsection{Discussion}

With this new projection formulation, one can formulate the QG system using the same prognostic variables as the RSW system, namely the horizontal velocity $(\bu, \bv)$ and the layer thickness $\bh$. This restores the proximity between these two equation sets in the ocean model hierarchy. We see four practical implications. First, given a RSW discretization, one has simply to implement the projection to have the companion QG model of this particular discretization. Second,  this formulation gives  access to the ageostrophic velocity hidden in the QG equations at the cost of a simple subtraction.
This allows an \textit{a posteriori} diagnostic of the validity of the QG scaling.
For a given geostrophic state $(\bu_g, \bv_g, \bh)$ we can compute the corresponding ageostrophic velocity $(\bu_a, \bv_a)$ and check that
\begin{equation}
 \frac{\bk_a}{\bk_g} \sim Ro^2 \,,\qquad
 \frac{\bh}{\bH} \sim Ro \ ,
 \label{eq:qgscaling}
\end{equation}
where $\bk_a=(|\bu_a|^2 + |\bv_a|^2)/2$ and $\bk_g=(|\bu_g|^2 + |\bv_g|^2)/2$ are respectively the ageostrophic and the geostrophic kinetic energies.
When this is  is not the case, the QG approximation is not valid and the solution produced by the QG model becomes questionable. 
Third, using the same variables $(\bu,\bv,\bh)$ opens the route to formulate eddy parameterizations, \textit{e.g} \cite{bachman2019gm+,li2023stochastic}, that work similarly on the RSW and the QG models, which is not the case when different formulations and/or discretizations are used.
Four, this formulation might give new insights and approaches for mathematical analysis of the QG equations.
One could look at the extension of recent results on the well-posedness properties for a stochastic RSW model \citep{crisan2023well} by studying the effect of the QG projector on these properties.

Finally, this formulation presents an interesting analogy with the Leray projector formulation of the Navier-Stokes equations.
The Leray projector enforces the incompressible constraint by filtering out the compressible sound waves.
It involves solving an elliptic Poisson equation for the pressure.
The QG projector presented here enforces the QG balance by filtering out the gravity waves.
It involves solving an elliptic Helmholtz equation for the PV. It worth emphasizing that the QG balance is not a constraint added to the RSW model. Indeed, the Hamiltonian structure of the QG model \citep{HolmZeitlin1998} is very different from the RSW one. In particular, and contrary to the pressure,  the PV is not a Lagrange multiplier enforcing a constraint.

\section{Numerical experiments}

\subsection{Numerical implementation}
To implement our method, we  need a RSW solver and a QG projector.
\subsubsection{RSW discretization}
We adopt the RSW solver\footnote{Available at https://github.com/pvthinker/pyRSW .} developed by \cite{roullet2022fast}, that we re-implemented in PyTorch for seamless GPU acceleration. We have verified that our implementation enables reproducing their original results up to numerical precision. The key element of the discretization is a fifth-order WENO upwinded reconstructions on both the the mass flux and the nonlinear vortex-force term. This provides enough mixing and dissipation, removing the need of an ad-hoc hyperviscous diffusion, while ensuring good material conservation of PV \citep{roullet2022fast}.

\subsubsection{QG discretization}
The QG projector requires an elliptic solver.
\cite{thiry2023MQ} released an efficient Python-PyTorch implementation of the QG equations.
They solve the QG elliptic equation with discrete sine transform implemented with PyTorch FFT, which uses highly optimized MKL FFT on Intel CPUs and cuFFT on Nvidia GPUs.
We use their elliptic solver to solve the QG elliptic equation \eqref{eq:qgmell}. Once the projector is available, the QG solver is implemented by simply adding the projection step to the RSW solver. Numerically the two solvers differ by one line of code, the projection step. All the other ingredients: variable staggering, time discretization, core RSW equations are identitical.

As a result, we obtain a PyTorch implementation\footnote{Will be released at publication time.} which is concise ($\sim$ 800 lines of code), as close as possible to the equations and which implements both the multi-layer QG and the RSW equations with the same state variables $(\bu, \bv, \bh)$.

\subsection{Vortex shear instability}
\begin{figure}
    \centering
    \includegraphics[width=0.4\linewidth]{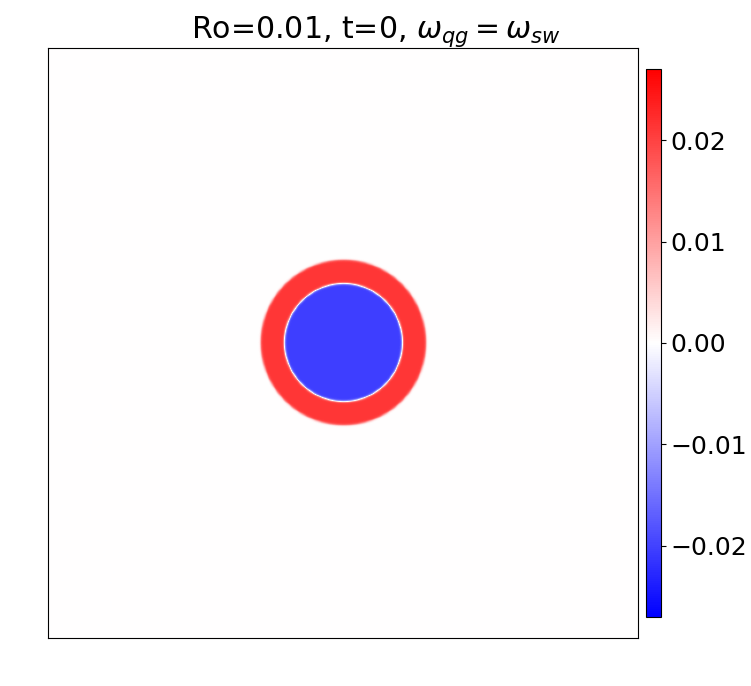}
    \includegraphics[width=0.4\linewidth]{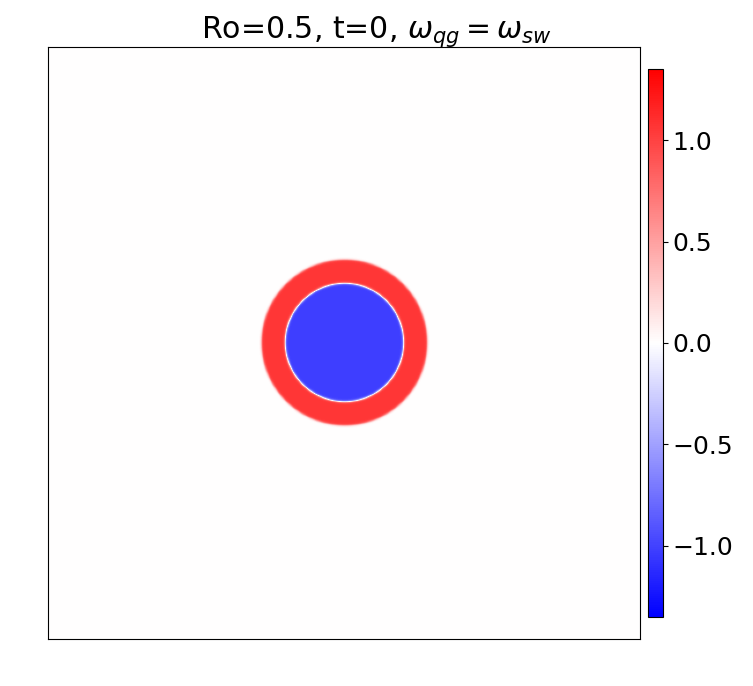}
    \includegraphics[width=0.95\linewidth]{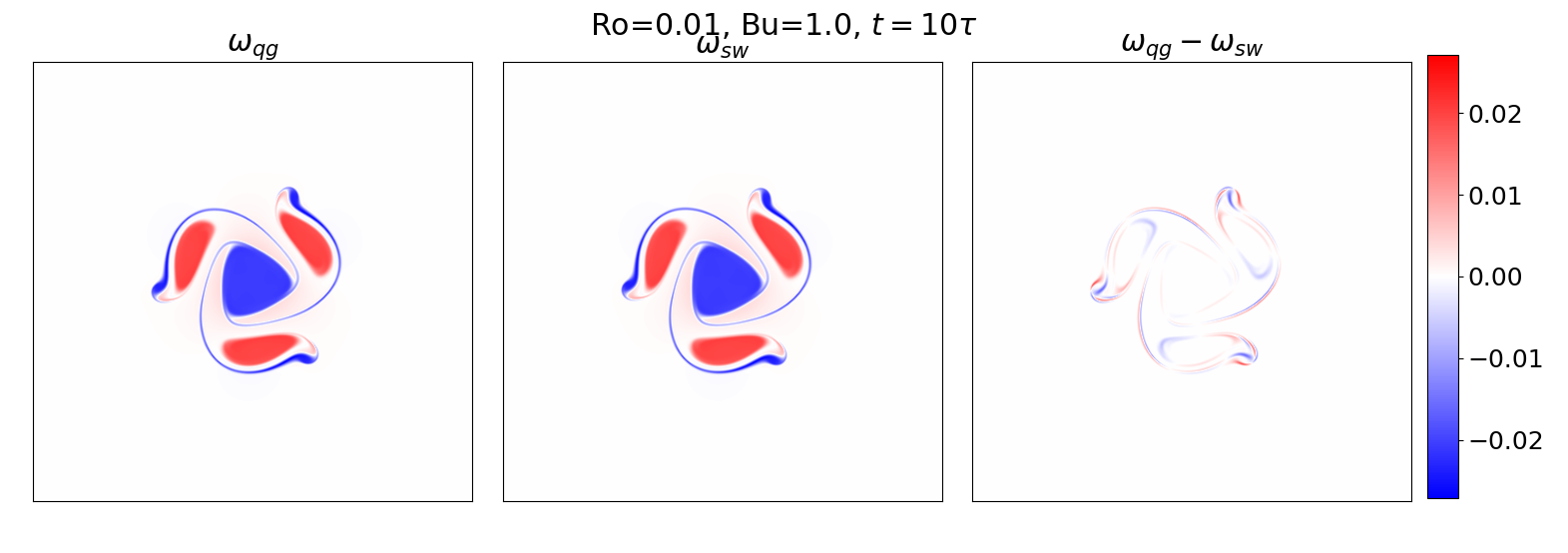}
    \includegraphics[width=0.95\linewidth]{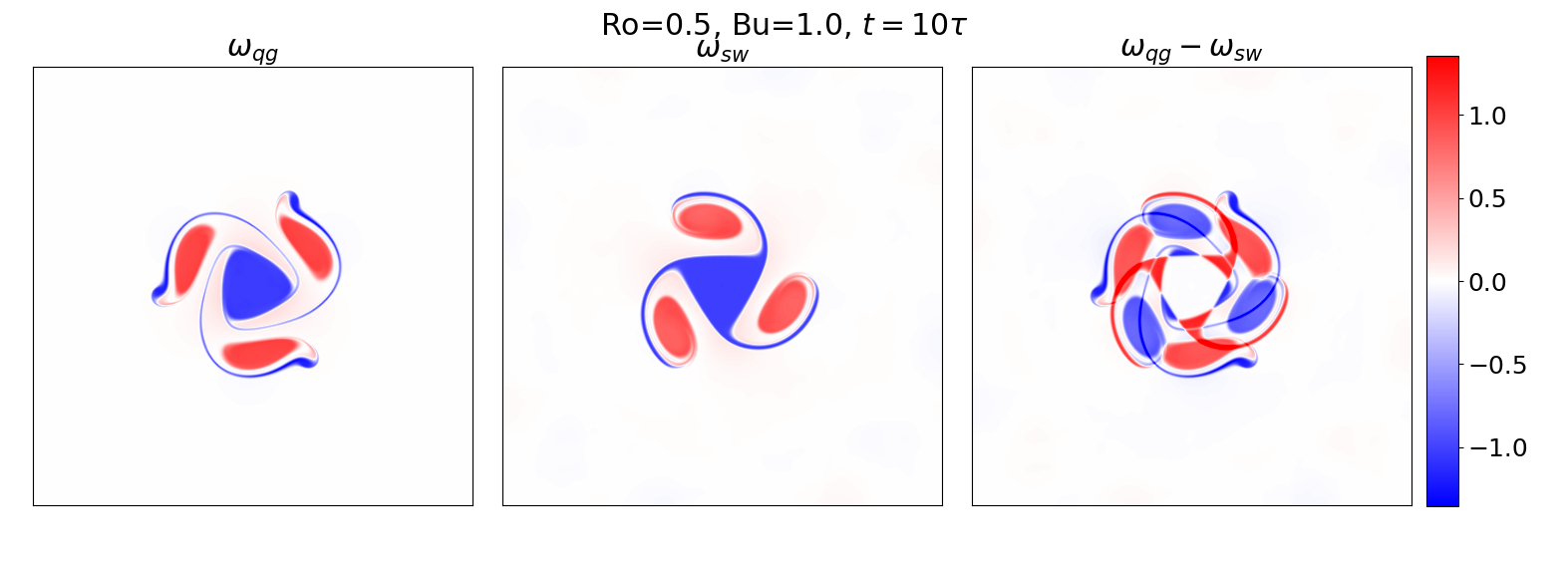}
    \caption{
      Vortex shear instability solved with QG and SW, $Bu=1$.
      (Top to bottom) Initial relative vorticity, $Ro=0.01$ and $Ro=0.5$.
      $Ro=0.01$, QG and SW final relative vorticities and difference.
      $Ro=0.5$, QG and SW final relative vorticities and difference.
      Unit $s^{-1}$.
    }
    \label{fig:vortex_plot}
\end{figure}

To validate our new formulation, we first study a vortex shear instability and compare its evolution in both the QG and the RSW models as $Ro$ increases for $Bu=1$. The initial state is a perfectly shielded vortex composed of a core of uniform vorticity $\omega_1$ surrounded with a ring of opposite sign uniform vorticity $\omega_2$. The system involves two lengths: the core radius $r_0$ and the vortex outer radius $r_1$. The ratio $\omega_2/\omega_1$ is such that the total circulation vanishes. This system is unstable and leads to multipoles formation \citep{morel1994multipolar}. The number of poles depends on the ratio of the vortex radius to the core radius.
We focus here on a tripole formation case, viz. $r_1/r_0=1.4$. To promote the growth of the most unstable mode we add a small mode 3 azymuthal perturbation. The experiments are set up in dimensional form with a square domain of size $L_x \times L_y =100\,\text{km}\times 100\,\text{km}$ on a f-plane. There is a single layer of fluid with thickness $H = 1\,\text{km}$. We assume no-flow and free-slip boundary conditions. The acceleration of gravity is set to $g = 10\,\text{m}\,\text{s}^{-2}$. The vortex core has a radius $r_0 = 10\,\text{km}$
and a positive vorticity, the vortex radius is $r_1=14\,\text{km}$.
We run the simulations on a $512\times 512$ grid, \textit{i.e.} 200m spatial resolution.
The Coriolis parameter is chosen such that $Bu=1$, viz. $f_0=\sqrt{gH}/r_0$. The vorticity $\omega_1$ is set indirectly via $u_{\rm max}$, the maximum velocity of the initial condition. We define the Rossby number as $Ro = u_{\rm max}/(f_0r_0)$. We impose $Ro$ and deduce $u_{\rm max}$. We have tested four cases: $Ro\in\{0.01,\, 0.05,\, 0.1,\,0.5\}$. In all cases we apply the QG projector to the initial state prior starting the time integration. The consequences are that the initial states are in geostrophic balance and differ only by a scaling factor. As $Ro$ increases the instability develops faster, in dimensional time. Therefore we present the results in the rescaled eddy-turnover time $\tau$, defined as the inverse of the $\ell^2$-norm of the initial vorticity, \textit{i.e.} $\tau = 1/\| \omega\|$. We integrate the simulations for the time $T = 10\, \tau $.

\begin{figure}
    \centering
    \includegraphics[width=0.99\linewidth]{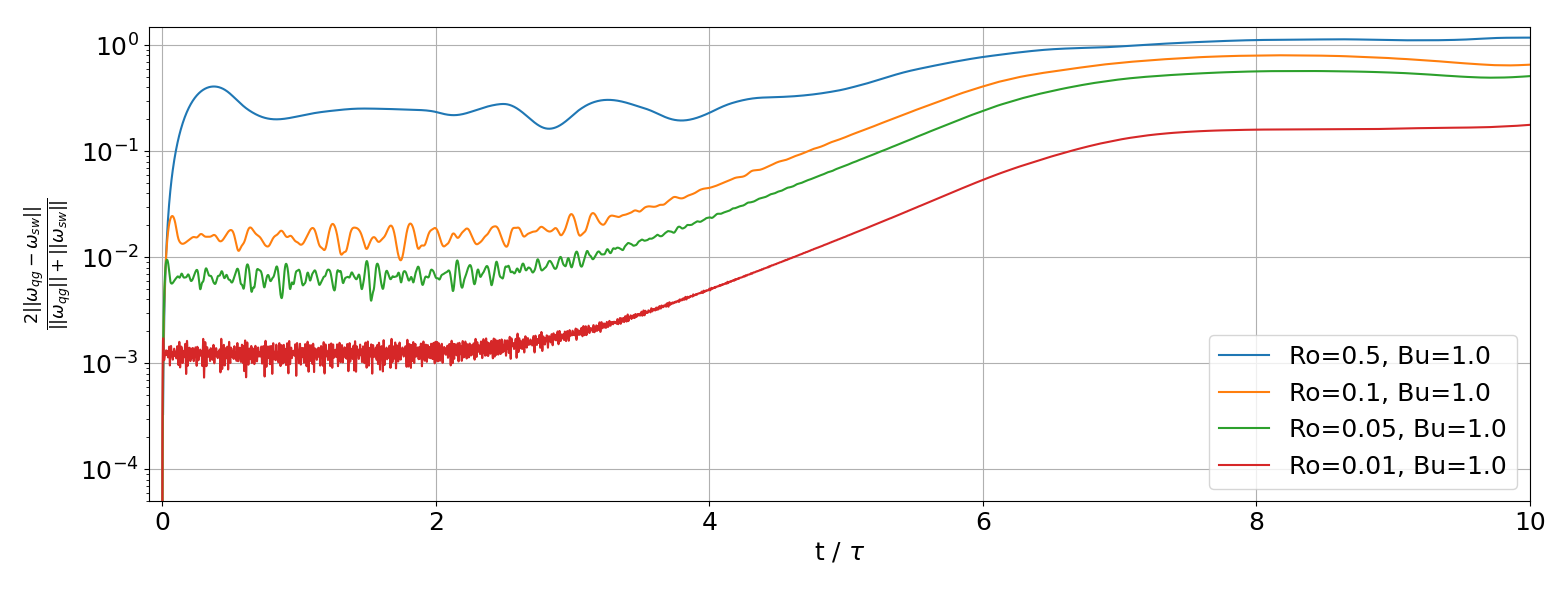}
    \caption{Time evolution of the normalized differences of relative vorticity between the QG model and the RSW model with $Bu=1$ and different $Ro$.
    }
    \label{fig:vortex_diff}
\end{figure}

Figure~\ref{fig:vortex_plot} shows how the initial state has evolved for the two extreme cases $Ro=0.01$ and $Ro=0.5$. In the $Ro=0.01$ case the differences between the QG and the RSW solutions are not perceptible to the naked eyes. This is consistent with the QG scaling and the fact that the QG model is an asymptotic limit of the RSW model. In the $Ro=0.5$ case the differences are of order one. In the RSW model the vortex spins faster and the filaments of negative vorticity are stabilized. The smoothness of the RSW solution at $t=10\,\tau$ might be surprising as we expect a certain amount of gravity waves generated during the fast cyclo-geostrophic adjustment of the initial state. These gravity waves are well present but at $t=10\,\tau$ they have bounced back and forth several times along the boundary and are strongly scattered. Interestingly the QG solution is exactly the same than in the $Ro=0.01$ case. This is expected since the QG equations are scaling invariant, viz. the evolution is invariant under the multiplication by a constant. However this is also quite remarkable to recover this property in the numerical solutions because the QG solver relies on the full RSW right-hand side. Another symmetry of the QG model is the parity invariance. The solution should be invariant under a sign change of the vorticity. In other words, cyclones and anticyclones follow the same evolution, up to a change of rotation. This is not at all the case for the RSW model. By flipping the sign of the initial vorticity we have checked that the solutions do satisfy this property (not shown). This means that the QG projector behaves well as it restores two invariances, scaling and parity, that are absent in the RSW solver. 

To assess the influence of $Ro$ on the time evolution, we define the normalized difference
$  \delta = 2||\omega_{\rm qg} - \omega_{\rm sw}||/(||\omega_{\rm qg}|| + ||\omega_{\rm sw}||)$.

Figure~\ref{fig:vortex_diff} shows $\delta(t)$ for the four $Ro$ cases. The oscillations during the $[0,\,4\,\tau]$ period are due to the gravity waves in the RSW case that result from the imperfect initial balance. Interestingly, the shortening of the oscillation period as $Ro$ decreases is due to the time rescaling by $\tau$. In dimensional time, the periods are the same. A pratical consequence is that the RSW experiment requires more time steps to reach $t=10\,\tau$ as $Ro$ decreases, whereas for the QG experiment the number of time steps is constant. After $t=4\,\tau$ the differences are dominated by the shear instability developing on the vortex. The small difference seen in the snapshots in the $Ro=0.01$ case have actually reached a plateau, meaning the QG solution closely follows the RSW one. This confirms that the QG model is a good simplified model when the scaling assumption holds. On a longer time scale and with a chaotic vortex dynamics, the solutions would diverge but on such a simple setup, the solutions remain close. The time evolution of $\delta$ is similar for all $Ro$ except $Ro=0.5$, because $\delta$ saturates at one, the maximum value of this metric. In this case, this metric suggests that two models predict a completely different solution. Looking back at the snapshots, this seems exagerated. The order one difference is mostly due to the difference in the timing, not to the difference in pattern. Depending on the purpose, the QG solution might still be of interest as it still captures the main phenomenology. Note that the QG solution could be made closer to the RSW solution if the two models were started with a different initial state: a cyclo-geostrophic balance for the RSW model and the associated projected state for the QG model. The initial state would  thus depend on $Ro$. For this illustrative experiment we have prefered to stick to the same initial state, up to a scaling constant.

\subsection{Double-gyre configuration}

To explore a richer phenomenology we have tested our new formulation on a classical oceanic test case, the idealized double-gyre. The domain is a non-periodic square ocean basin with $N=3$ layers. We assume free-slip boundary conditions on each boundary.
We apply a stationary and symmetric wind stress $ (\tau_x, \tau_y)$ with $\tau_x = - (\tau_0 / \rho_0) \cos(2 \pi y / L_y)$ and $\tau_y = 0$ in  the top layer and a linear drag in the bottom with drag coefficient $\gamma$.
The parameter values are given in Table \ref{tab:params} in Appendix A.

We study this configuration in an eddy-permitting resolution of 20~km, meaning that the spatial resolution (20~km) is half of the largest baroclinic Rossby radius (41~km).
We expect QG and RSW simulations to produce strong western boundary currents converging to the middle of the western boundary and an eastward jet departing from the middle of the western boundary.
Starting from the rest state, this configuration requires about 100 years to spin-up and have converged statistics \citep{hogg2005mechanisms,simonnet2005quantization}.

\begin{figure}
    \centering
    \includegraphics[width=0.99\linewidth]{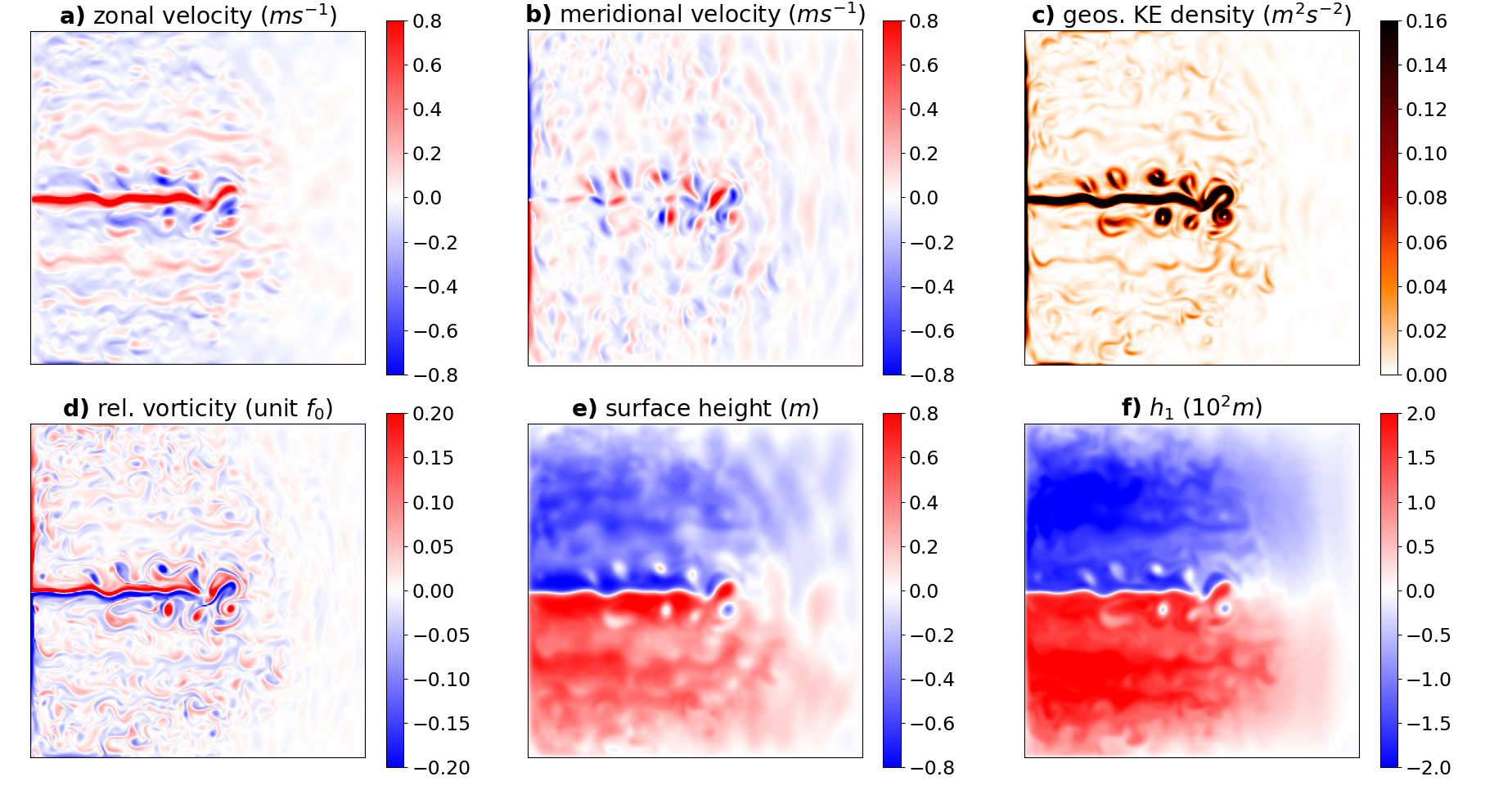}
    \caption{
        QG upper-layer snapshot after 100 years of spin-up from rest state.
        KE, geos., rel. stand respectively for kinetic energy, geostrophic and relative.
    }
    \label{fig:qg_snapshot}
\end{figure}
In Figure~\ref{fig:qg_snapshot} we present snapshots of key quantities of the upper layer after 100 years of integration. The solution exhibits the expected properties of this kind of setup: a double-gyre system (Fig.~\ref{fig:qg_snapshot}e) separated by a meandering eastward jet (Fig.~\ref{fig:qg_snapshot}a,d) emanating from the western boundary, two strong and narrow western boundary currents (Fig.~\ref{fig:qg_snapshot}b) feeding the eastward jet, mesoscale turbulence everywhere and intensified near the jet (Fig.~\ref{fig:qg_snapshot}c,d), Rossby waves propagation (Fig.~\ref{fig:qg_snapshot}e,f). The local Rossby number defined as $\omega/f_0$ peaks at 0.2 which is not small but not large enough to dismiss the solution either. Interestingly the layer thickness perturbation $h_1$ is not small at all compared to the reference depth $H_1=400\,\text{m}$.

\begin{figure}
    \centering
    \includegraphics[width=0.99\linewidth]{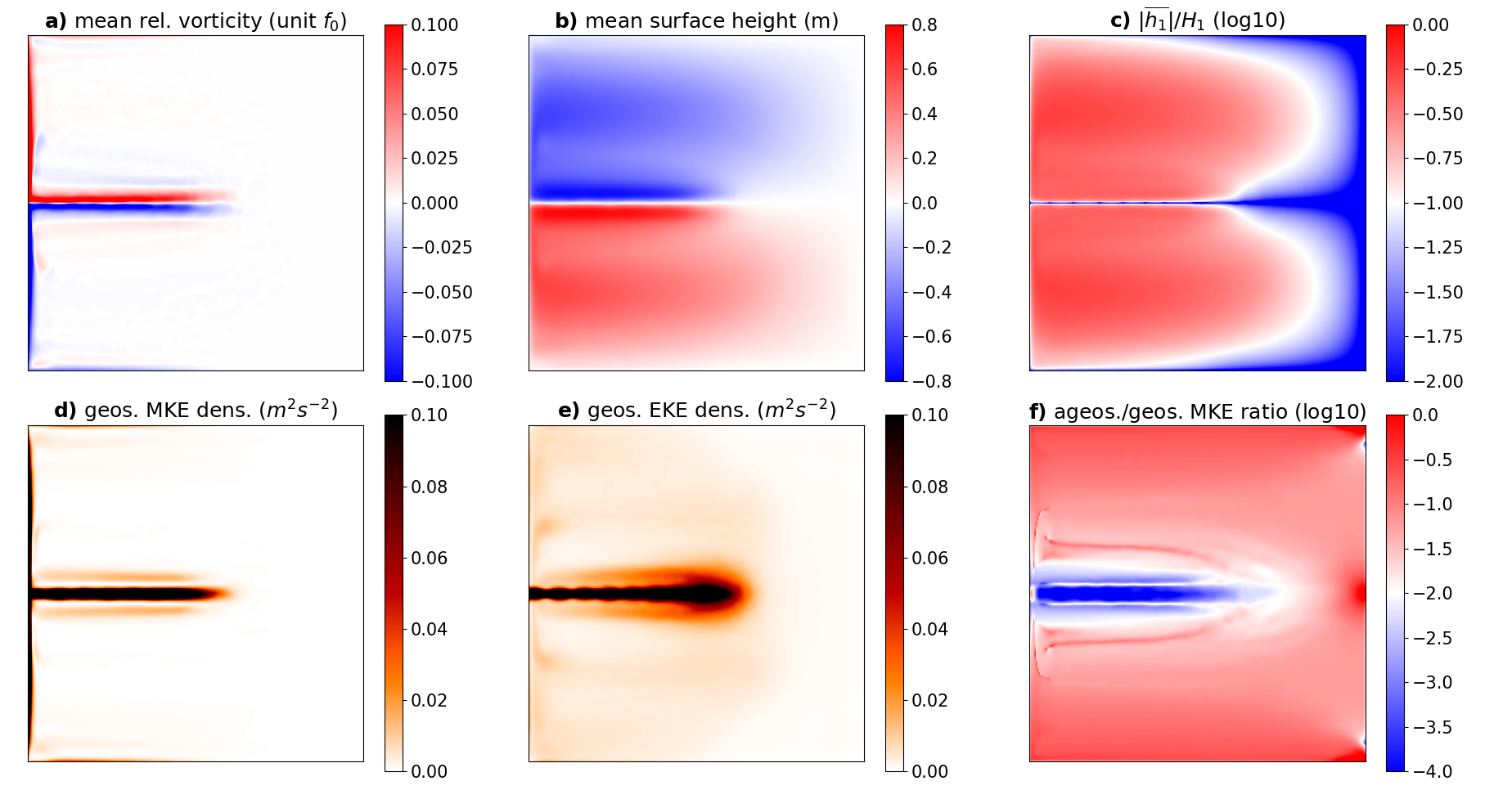}
    \caption{QG upper-layer statistics over 100 years after 100 years spin-up from rest state.
    }
    \label{fig:qg_stats}
\end{figure}
To compliment this one snapshot, we present statistics estimated over the years [100-200] after the initial state and using one snapshot every 10 days. We decompose the geostrophic kinetic energy into its mean and its eddy part. We also compute the $k_a/k_g$ ratio. The results are shown in Figure~\ref{fig:qg_stats} for the upper layer.
The statistics confirm the presence of the strong western boundary currents (Fig.~\ref{fig:qg_stats}a) as well as the strong eastward jet reaching the middle of the domain (Fig.~\ref{fig:qg_stats}a,d). A remarkable feature of the solution is the symmetry of all the quantities with respect to the central latitude. This is due to both the symmetry of the forcing and the fact that the QG model has a cyclone/anticyclone symmetry which prevents the model to break the symmetry of the forcing. From the ratio $k_a/k_g$ and $h/H$ (Fig.~\ref{fig:qg_stats}c,f) we can test the validity of the QG scaling assumptions (Eq.~\ref{eq:qgscaling}). The colorscale has been adjusted so that the white intermediate color corresponds to $Ro=0.1$. The red areas indicate where the QG scaling assumptions are not respected. Counter-intuively the central jet region, where the kinetic energy is the largest, is where the QG scaling is the best. The worst regions are the gyres centers because of too large thickness deviations, and along the boundaries because of too large ageostrophic velocities.

\begin{figure}
    \centering
    \includegraphics[width=0.99\linewidth]{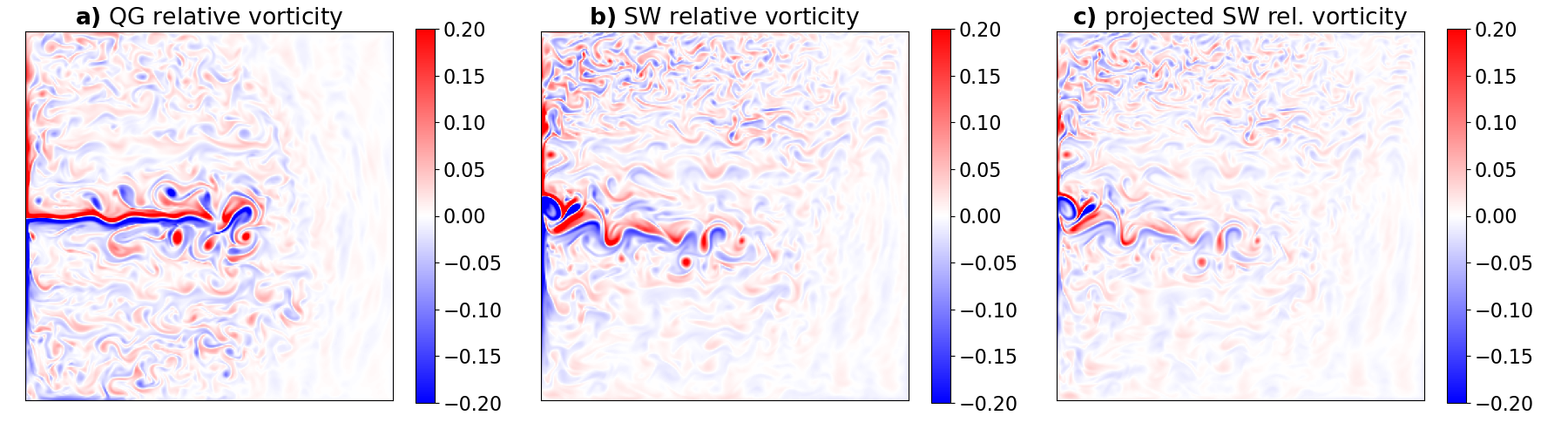}
    \caption{(Left) QG upper-layer relative vorticity after 100 years spin-up, (middle) RSW upper-layer relative vorticity and (right) velocity divergence after 2 years additional spin-up.
    Unit of $f_0$.
    }
    \label{fig:qg_sw_snapshot}
\end{figure}
Finally, we compare the QG solution with its companion RSW one. But, because of the free surface, and the presence of fast barotropic gravity waves, the RSW solver requires a much smaller time step, typically 200 times smaller. This 200 factor corresponds to the ratio $\sqrt{gH}/\max(u)$. Integrating from the rest state over 200 years would be a waste of computational ressources. A compromise would be to use a  barotropic-baroclinic time splitting \citep{higdon1997barotropic} or an implicit stepping of the free-surface \citep{roullet2000salt} but this goes beyond the scope of the present study. Instead, we ran the RSW solver starting from the year 200 of the QG solution, and we integrate it over 2 years. Figure~\ref{fig:qg_sw_snapshot} compares QG and RSW on a snapshot of vorticity in the upper layer. The central jet has now a southward component. The two gyres are no longer symmetric. The mesoscale turbulence has intensified in the Northern gyre and weakened in the Southern gyre. Interestingly if we apply the QG projector on the RSW state the resulting state remains very close (Fig.~\ref{fig:qg_sw_snapshot}c) albeit with damped fluctuations. The QG projector tends to dampen the short scales but, since it is applied on the RSW model tendency, it does not affect the QG state, which does not present this dampening as we can check from a visual comparison between Fig.~\ref{fig:qg_sw_snapshot}a and Fig.~\ref{fig:qg_sw_snapshot}b. The closeness suggests that while the QG solution might be locally tangent to the RSW solution, when integrated over a long time, the differences build up and yield a quite different state.

\section{Conclusion}

In this paper we have shown that the QG model can be formulated as a projected RSW model, namely by applying the QG projector $P$ (defined in Equation \ref{eq:qg_projector}) to the RSW tendency. This allows to manipulate a  QG model with the same variables $(\bu,\,\bv,\,\bh)$ than a RSW model, instead of the more usual $(\bp,\,\bq)$ variables. This helps better understand the proximity and the differences between the two models. It also allows to have a numerical model capable of integrating the two models with the same numerics. We have tested on a vortex shear instability that the resulting QG model has all the desired properties of symmetry: scaling and parity, that are absent in the RSW model. We have shown that a QG model implemented in this way recovers the expected features of a double-gyre experiment. A direct consequence of the approach is the capability to compare the RSW and the QG solutions.

This new formulation opens the path for studies of differences between QG and RSW equations, for example study the stability of the geostrophic equilibrium or the study of spontaneous imbalance.
It might be very beneficial for data assimilation applications as it allows to easily switch between RSW and QG models, or even to combine them.
One could for example run a first data assimilation algorithm, \textit{e.g.} ensemble Kalman filter \citep{evensen2003ensemble}, using the projected QG to capture the geostrophic dynamics, and then to refine the obtained solution by switching to the RSW model.

\backsection[Acknowledgements]{
  The authors acknowledge the support of the ERC EU project 856408-STUOD.
  The authors would like to thank Anne-Laure Dalibard for pointing them to Frederic Charve's Thesis, and Frederic Charve for interesting discussion.
}

\backsection[Funding]{ERC EU project 856408-STUOD. OcéanIA INRIA Challenge project}

\backsection[Declaration of interests]{The authors report no conflict of interest.}

\backsection[Data availability statement]{Code to reproduce results will be released at publication time.}


\backsection[Author contributions]{
    Original idea by GR.
    LT, LL, and GR derived the math.
    LT and LL implemented the software supervised by GR.
    LT ran the numerical experiments and produced the figures.
    LT and GR wrote the article.
    LL, and EM helped the redaction with suggestions and corrections.
    EM provided the funding.
}

\appendix

\section{Double-gyre configuration}
\label{app:A}

We provide the physical and numerical parameters used in the double-gyre configuration in Table \ref{tab:params}.
\begin{table}
\begin{center}
\def~{\hphantom{0}}
\begin{tabular}{c|c|c}
Parameter   & Value                              & Description           \\\hline
$L_x, L_y$  & $5120, 5120$ km                    & Domain size               \\
$H_k$       & $400, 1100, 2600$ m                & Mean layer thickness   \\
$g_k'$      & $9.81, 0.025, 0.0125$ m s$^{-2}$   & Reduced gravity       \\
$\gamma$    & $3.6\ 10^{-8}$ s$^{-1}$            & Bottom drag coefficient   \\
$\tau_0$    & $0.08$ N m$^{-2}$                  & Wind stress magnitude \\
$\rho$      & $1000$ kg m$^{-3}$                 & Ocean density               \\
$f_0$       & $9.375\ 10^{-5}$ s$^{-1}$          & Mean Coriolis parameter   \\
$\beta$     & $1.754\ 10^{-11}$ (m s)$^{-1}$     & Coriolis parameter gradient   \\
$L_d$       & $41, 25$ km                        & Baroclinic Rossby radii   \\
$n_x, n_y$  & $256, 256$                         & Grid size\\
$dt$        & 4000 s                             & Time-step\\
\end{tabular}
\caption{Parameters of the idealized double-gyre configuration}
\label{tab:params}
\end{center}
\end{table}
There no viscosity nor diffusion coefficient as the grid-scale dissipation is implicitely handled by the upwinded WENO reconstruction of the masse flux and the vortex-force \citep[see][]{roullet2022fast}.

\clearpage

\bibliographystyle{jfm}
\bibliography{jfm_GR}

\end{document}